\documentclass[]{spie}  

\usepackage{amsmath,amsfonts,amssymb}
\usepackage{graphicx}
\usepackage[colorlinks=true, allcolors=blue]{hyperref}
\usepackage{multirow}
\usepackage{array}
\usepackage{longtable}
\usepackage{soul}
\usepackage{enumitem}

\usepackage{xcolor,colortbl}
\usepackage[]{aas_macros}

\graphicspath{{figures/}}

\title{Quantum-optimal instruments for high-contrast imaging with multi-plane light converters}

\author{
Sebastiaan Y. Haffert$^{*,a,b}$,
Yinzi Xin$^{a}$,
Rico Landman$^{a}$
\\
\vspace{0.2cm}
$^{a}$Leiden University, Einsteinweg 55, 2233 CC, Leiden, The Netherlands\\
$^{b}$Steward Observatory, 933 North Cherry Avenue, Tucson, AZ 85721-0065, USA}

\authorinfo{*sebastiaan.haffert@strw.leidenuniv.nl}

\begin{document} 
\maketitle


\begin{abstract}Spatial Mode Demultiplexers (SPADE) are devices that split an incoming wavefront into a particular set of electric field modes. SPADEs have been shown to be quantum-optimal for many measurement problems including coronagraphy, wavefront sensing and interferometric (nulling) beam combiners. We will show how quantum optimal SPADEs can be implemented using Multi Plane Light Converters (MPLC). MPLCs are devices that use several phase plates in sequence to achieve arbitrary unitary transformations. The phase plates themselves are implemented using liquid-crystal technology that allows us to create high-quality broadband phase plates. The accurate control of the local phase through the direct write method results in MPLCs with minimal background scatter. We will show the designs and simulated performance for each of the proposed use cases for the ELTs, HWO and LIFE and show how they improve over current concepts.\end{abstract}



\section{Introduction}

One of the most technically challenging tasks in optical system design is the direct detection and characterization of biosignatures in the atmospheres of exoplanets, planets outside our own Solar System (see Figure 1). Three upcoming observatories are specifically designed to enable the search for signs of life in the Universe:

\begin{itemize}
    \item The Giant Segmented Mirror Telescopes (GSMTs), including the European Southern Observatory (ESO) Extremely Large Telescope (ELT) and the Giant Magellan Telescope (GMT), will offer unprecedented sensitivity and spatial resolution thanks to their enormous apertures. These capabilities will enable characterization of exoplanets at the smallest angular separations \cite{Kasper2021PCS}.
    
    \item NASA's Habitable Worlds Observatory (HWO) is a large ultraviolet/optical/infrared space telescope designed specifically to search for signs of life on planets orbiting nearby stars. The mission is currently planned for launch in the 2040s \cite{Feinberg2026HWO}.
    
    \item The Large Interferometer For Exoplanets (LIFE) is a space-based mission concept consisting of four telescopes operating in a formation-flying configuration to create a mid-infrared interferometer capable of characterizing exoplanet atmospheres \cite{Quanz2022LIFE}.
\end{itemize}

These observatories provide complementary capabilities that together will enable detailed characterization of nearby exoplanet atmospheres and searches for remote biosignatures. Such measurements could address one of humanity's oldest questions: are we alone in the Universe?

However, realizing this ambitious goal requires new technologies that push astronomical instrumentation toward the fundamental physical limits of measurement. Current high-contrast imaging systems remain limited by technology rather than fundamental physics, and significant improvements beyond existing approaches may still be possible \cite{Belikov2021CoronagraphLimits}. The coming years, before the next-generation observatories finalize their technology selections, therefore represent a critical opportunity to develop transformative instrumentation.

In this proceeding, we propose the development of quantum-enhanced optical instruments designed to approach the fundamental information limits imposed by quantum mechanics under realistic astronomical observing conditions.

\section{Quantum Optimal Measurements}

Quantum metrology is a rapidly growing field that exploits the principles of quantum mechanics to achieve the highest possible precision in parameter estimation. The ultimate precision limit of a measurement is set by the Quantum Cramér--Rao Bound (QCRB), which states that the variance of an unbiased estimator is lower bounded by the inverse of the Quantum Fisher Information (QFI) \cite{Braunstein1994QFI}. A system with a high QFI contains more information about the parameter being estimated and therefore enables measurements with lower uncertainty.

The QFI depends on both the quantum state of the system and the parameters being estimated. Quantum metrology provides the tools to determine this fundamental information content, thereby establishing the maximum achievable precision of a measurement. However, the QFI alone does not specify how this information can be extracted experimentally, as it is optimized over the complete space of possible measurements.

In contrast, the Classical Fisher Information (CFI) describes the information extracted by a specific measurement implementation. A measurement is quantum-optimal when its CFI reaches the QFI bound. The central challenge in quantum-enhanced instrumentation is therefore to design physical measurement architectures that can saturate these fundamental limits.

\subsection{Spatial Mode Demultiplexing (SPADE)}

Sorting collected photons into an optimized spatial mode basis prior to detection can significantly increase the information content of a measurement compared to conventional image-plane detection using pixelated detectors. These so-called Spatial-mode Demultiplexers (SPADEs) project the incoming optical field onto a set of orthogonal spatial modes that are subsequently detected independently using photon-counting detectors \cite{Tsang2016QuantumSuperresolution}.

For a single telescope, SPADE measurements have been shown to achieve the quantum Fisher information limit for estimating the separation of two incoherent point sources \cite{Tsang2016QuantumSuperresolution}. A remarkable property of SPADE is that the available information remains high even when the source separation is significantly smaller than the diffraction limit. This behavior is in stark contrast to conventional direct imaging, where the Fisher information rapidly decreases as the angular separation approaches zero, a phenomenon known as Rayleigh's curse \cite{Tsang2016QuantumSuperresolution}.

The sensitivity advantage of SPADE measurements persists in the high-contrast regime relevant for exoplanet imaging, where planets can be millions to billions of times fainter than their host stars \cite{deshler2026quantum}. Similar advantages have been demonstrated for more complex imaging problems, including localization of multiple point sources \cite{Tan2022RobustQuantumImaging}, estimation of object dimensions \cite{Dutton2019SpatialModeSorting}, and classification of objects from libraries \cite{Grace2022QuantumObjectIdentification}. More recently, quantum-optimal measurements have also been explored for discrete optical systems such as multi-telescope interferometers relevant for missions such as LIFE \cite{Padilla2026EntangledTelescopy}.

\subsection{Multi-Plane Light Converters (MPLC)}

Each astronomical measurement problem requires its own optimized spatial mode basis. An ideal SPADE architecture should therefore provide the flexibility to implement arbitrary mode transformations. However, many current implementations, such as mode-specific photonic lanterns, lack the required flexibility and scalability for next-generation large segmented telescopes.

Multi-Plane Light Converters (MPLCs) provide a promising solution. MPLCs use a sequence of spatially varying phase transformations separated by free-space propagation to implement high-dimensional optical transformations. It has been shown that arbitrary finite-dimensional unitary transformations can be implemented using sequences of phase operations and propagation steps \cite{Reck1994Unitary}. This mathematical framework forms the basis for the exceptional flexibility of MPLCs.

MPLC devices capable of multiplexing and demultiplexing hundreds of spatial modes have already been demonstrated \cite{Fontaine2017ModeMultiplexers,Fontaine2019LaguerreSorter}, significantly exceeding the modal complexity of many fiber-based approaches. However, widespread implementation of MPLCs remains limited by the fabrication of high-quality continuous phase elements.

Current implementations typically rely on spatial light modulators (SLMs) or multi-step lithography to realize phase transformations. These approaches introduce practical limitations, including restricted spectral bandwidth, polarization sensitivity, or finite phase precision due to discretization.

\subsection{Patterned Retarders with Liquid Crystal Technology}

Liquid crystals provide a powerful platform for geometric phase manipulation by controlling the spatial orientation of the optical axis through patterned birefringence. When polarized light propagates through a patterned liquid crystal element, spatial variations in the liquid crystal fast-axis orientation produce corresponding spatial variations in phase through the geometric, or Pancharatnam--Berry, phase \cite{Cohen2019GeometricPhase}.

Direct-write fabrication techniques enable precise control of liquid crystal alignment patterns with sub-micron spatial resolution \cite{Miskiewicz2014DirectWriteLC}. Furthermore, twisted multi-layer liquid crystal structures allow broadband achromatic phase control by combining multiple layers with carefully engineered retardance and twist profiles \cite{Komanduri2013MultiTwistRetarders}. Together, these techniques enable broadband geometric phase elements suitable for advanced astronomical instrumentation.

In Leiden, we have developed extensive expertise in applying geometric phase elements to astronomical instrumentation. These technologies have been demonstrated for coronagraphy \cite{Por2020SCAR,Doelman2021VectorApodizing,Doelman2025SUPPRESS}, wavefront sensing \cite{Doelman2018VectorZernike,Haffert2018LEXI,Roberts2022SIGHT}, phase mode demultiplexing, sparse aperture masking \cite{Doelman2021HolographicApertureMask}, astrometry \cite{Langford2024DiffractivePupil}, and precision phasing of the Giant Magellan Telescope \cite{Haffert2022GMTPhasing}.

Building on this unique expertise in geometric phase optics for astronomy, together with our collaboration with Colorlink Japan Ltd. and advanced inverse-design techniques \cite{Landman2022AutoDiffAO}, we aim to develop next-generation MPLC architectures capable of implementing quantum-optimal measurements for future exoplanet characterization missions.

\section{Example applications of MPLC-based SPADEs}

To demonstrate the flexibility of MPLCs as a platform for quantum-optimal astronomical instrumentation, we present three representative examples currently under development. Although each application addresses a different measurement problem, they are all based on the same physical principle: implementing an optimized unitary transformation that projects the incoming electric field onto the spatial modes that maximize the information content of the measurement. These examples illustrate that a single technological platform can be adapted to widely different astronomical instruments, ranging from single telescope imaging to space-based interferometry.

\subsection{Broadband mode matching for efficient fiber coupling}

\begin{figure}[ht]
\centering
\includegraphics[width=0.75\textwidth]{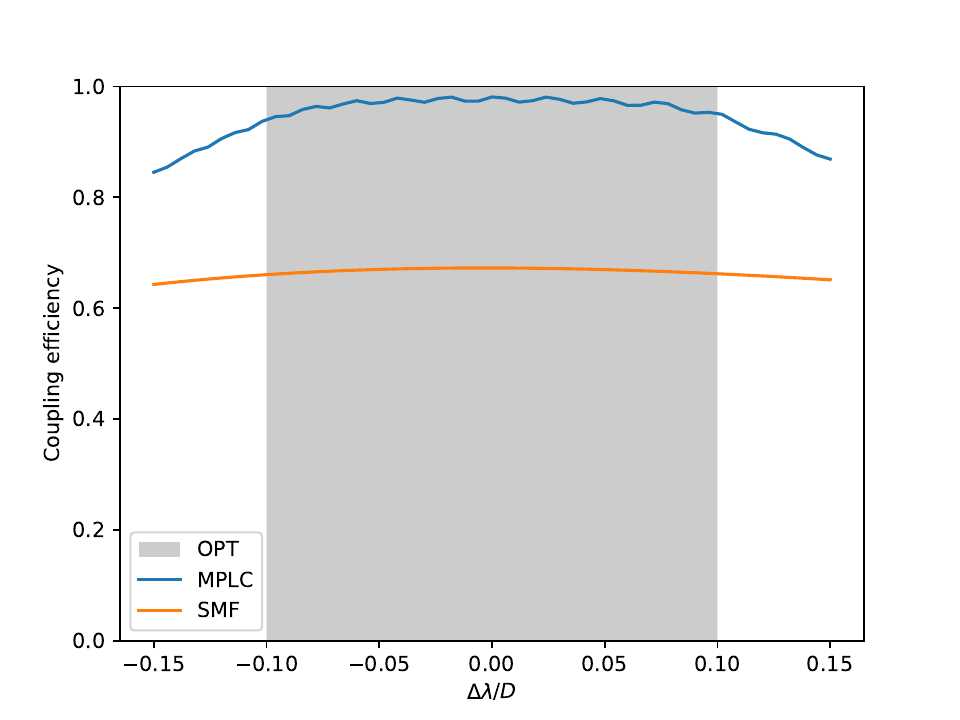}
\caption{Broadband ELT fiber coupling results. The coupling efficiency versus wavelength for a 5-layer MPLC design in orange and a conventional fiber in blue. The MPLC design outperforms a standard SMF over the optimized wavelength range (gray shaded area) and on the additional spectral range outside it.}
\label{fig:fiber}
\end{figure}

Single-mode fibers (SMFs) are fundamental building blocks of many astronomical instruments because they provide spatial filtering, stable beam transport and efficient interfaces to photonic devices. However, the coupling efficiency between a telescope pupil and the fundamental fiber mode is fundamentally limited by the pupil geometry\cite{shaklan1988coupling}. For centrally obscured and segmented apertures such as the ELT, the theoretical maximum coupling efficiency into a conventional Gaussian fiber mode is approximately 62\%.

Rather than modifying the fiber, an MPLC can be designed to perform the optimal unitary transformation between the telescope pupil and the guided mode of a standard single-mode fiber. The transformation redistributes the electric field while preserving optical throughput, allowing nearly all of the incoming light to be coupled into the fiber.

Figure~\ref{fig:fiber} shows the simulated performance of an MPLC designed for the ELT pupil. Across a spectral bandwidth of approximately 20\%, the coupling efficiency remains close to 100\%, substantially exceeding the conventional coupling limit. The broadband performance results from the achromatic geometric-phase implementation of the phase plates, while the low scattering produced by the direct-write fabrication process minimizes losses between successive planes.

This example demonstrates that MPLCs can effectively remove one of the long-standing efficiency limitations in astronomical fiber injection \cite{jovanovic2017efficient}. Such improvements directly increase the sensitivity of fiber-fed spectrographs, nulling interferometers and photonic processing units by increasing the number of detected photons without modifying the telescope optics.

\subsection{MPLC beam combiner for the LIFE mission}

\begin{figure}[ht]
\centering
\includegraphics[width=\textwidth]{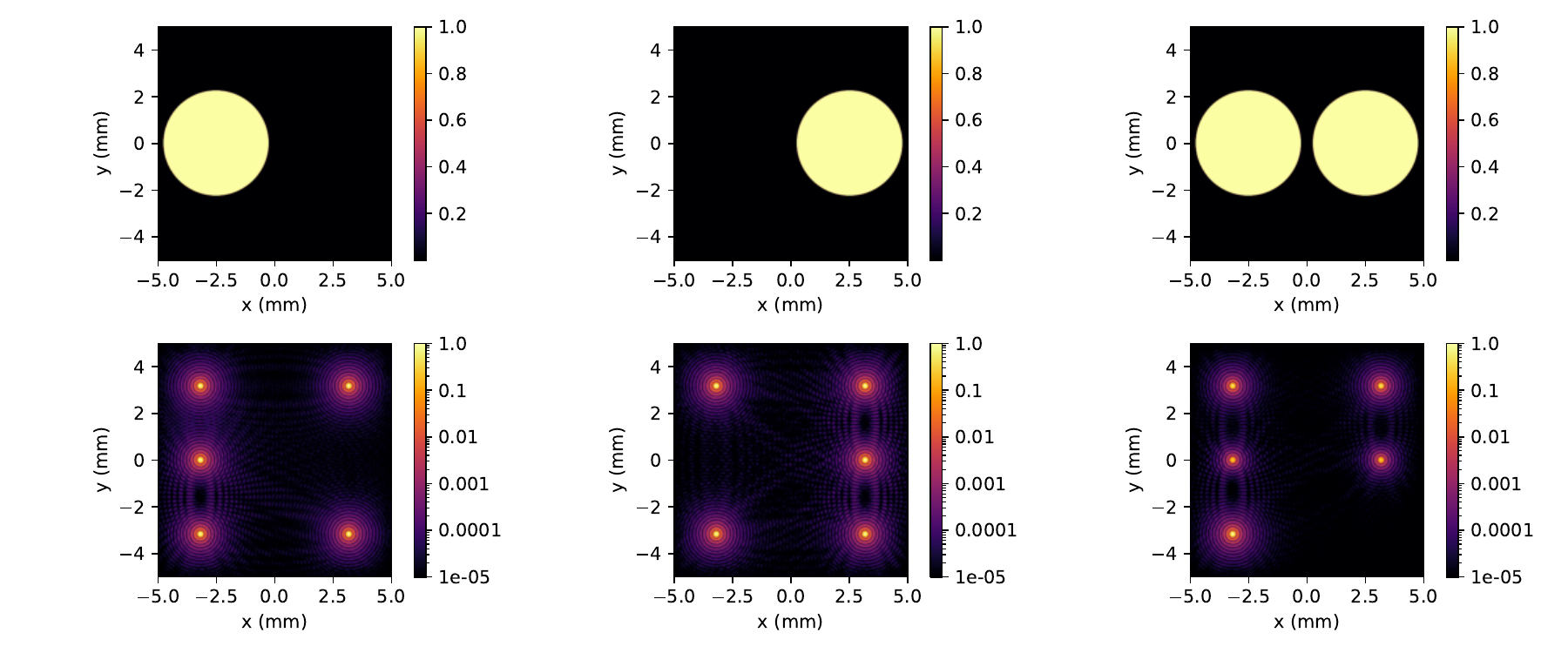}
\caption{The response of the MPLC for beam combining for the different sub-apertures and the combined apertures. The four PSFs in the corners are injected with different amounts of differential phase that build up the classic interferometric ABCD measurement. The spots in the middle are power monitoring spots and each aperture generates only one.}
\label{fig:life_apertures}
\end{figure}

The proposed LIFE mission relies on coherent beam combination from four free-flying telescopes to suppress stellar light while transmitting planetary emission\cite{quanz2022large}. The beam combiner therefore represents one of the key optical components of the instrument. Any leakage caused by imperfect mode orthogonality or cross-talk directly limits the achievable null depth\cite{hansen2023large}.

Because beam combination is fundamentally a unitary transformation, it can naturally be implemented using an MPLC. Rather than constructing the interferometric transfer matrix from multiple beam splitters and phase shifters, the complete transformation is synthesized through a sequence of optimized phase plates.

\begin{figure}[ht]
\centering
\includegraphics[width=\textwidth]{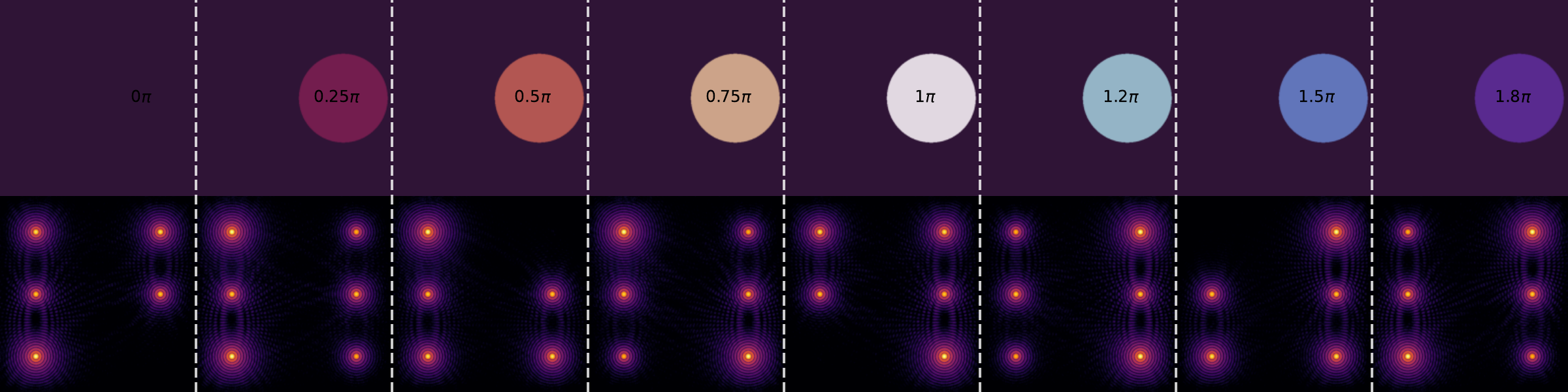}
\caption{The response of the MPLC-based beam combiner to various amounts of differential phase between the sub-apertures. The differential phase starts at zero phase on the left and increases to nearly 2$\pi$ on the right. The bottom row shows that the PSFs in the corner react to the differential piston.}
\label{fig:phase_shifts}
\end{figure}

Figure~\ref{fig:life_apertures} presents the simulated response of a two-telescope MPLC beam combiner demonstrating how it works. A more detailed response is shown in \ref{fig:phase_shifts} where various amounts of differential piston are injected. The optimized design achieves cross-talk levels between approximately $-60$ and $-70$~dB between the output channels, indicating extremely high modal purity. Such low cross-talk is particularly attractive for deep nulling interferometry, where leakage between the bright stellar outputs and the nulled outputs rapidly becomes the dominant error source.

Beyond its excellent optical performance, the MPLC implementation offers a compact monolithic architecture with no critical alignment between multiple beam splitters after fabrication. The same optimization framework can also be extended to larger interferometric arrays or beam combiners optimized for alternative science cases. It can also be combined with the single-mode filter from the previous section to create a single optical system that implements modal filtering and beam combining together. 

\subsection{MPLC-based SPADE coronagraph}
\begin{figure}[ht]
\includegraphics[width=0.5\textwidth]{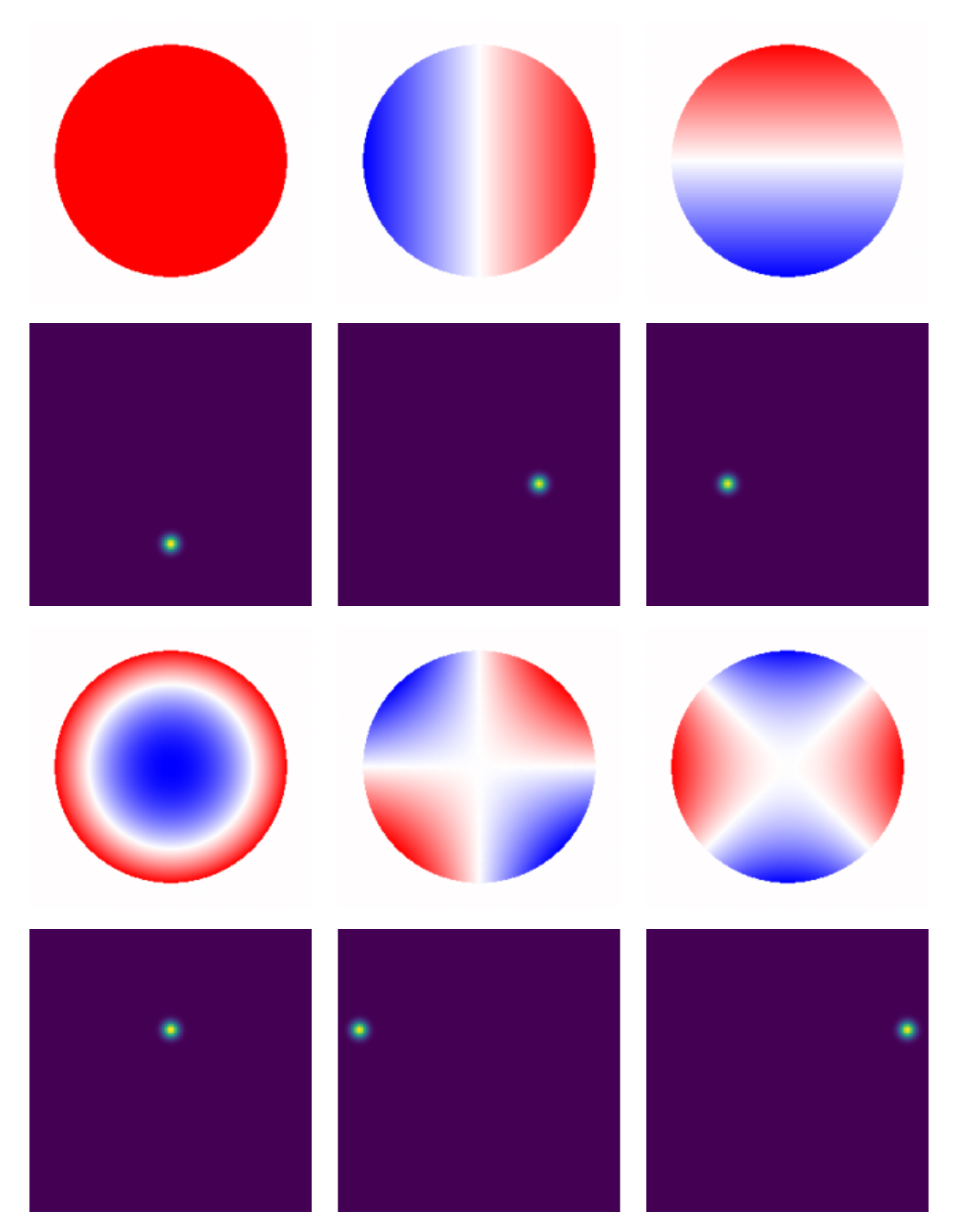}
\centering
\caption{The MPLC SPADE coronagraph sorting the first 6 modes into a hexagonal arangement. Each spot is shaped into a Gaussian spot to further reduce cross-talk caused by diffraction between the modes.}
\label{fig:coronagraph}
\end{figure}
Coronagraphy can also be interpreted as a spatial mode sorting problem\cite{deshler2026quantum}. Instead of minimizing the stellar intensity in the focal plane, the incoming electric field is projected onto an optimized modal basis that concentrates the stellar light into a small number of output channels while maximizing the distinguishability of faint off-axis companions.

An MPLC provides a natural platform for implementing such a quantum-optimal SPADE coronagraph because it can realize arbitrary spatial mode projections with high fidelity. The resulting device simultaneously performs the modal decomposition and routes each mode to an independent detector, allowing the full spatial information to be retained instead of being integrated over image pixels. This flexibility will be necessary to implement more complex mode sorting than is available with other technology platforms\cite{xin2026quantum}.

\begin{figure}[ht]
\includegraphics[width=0.5\textwidth]{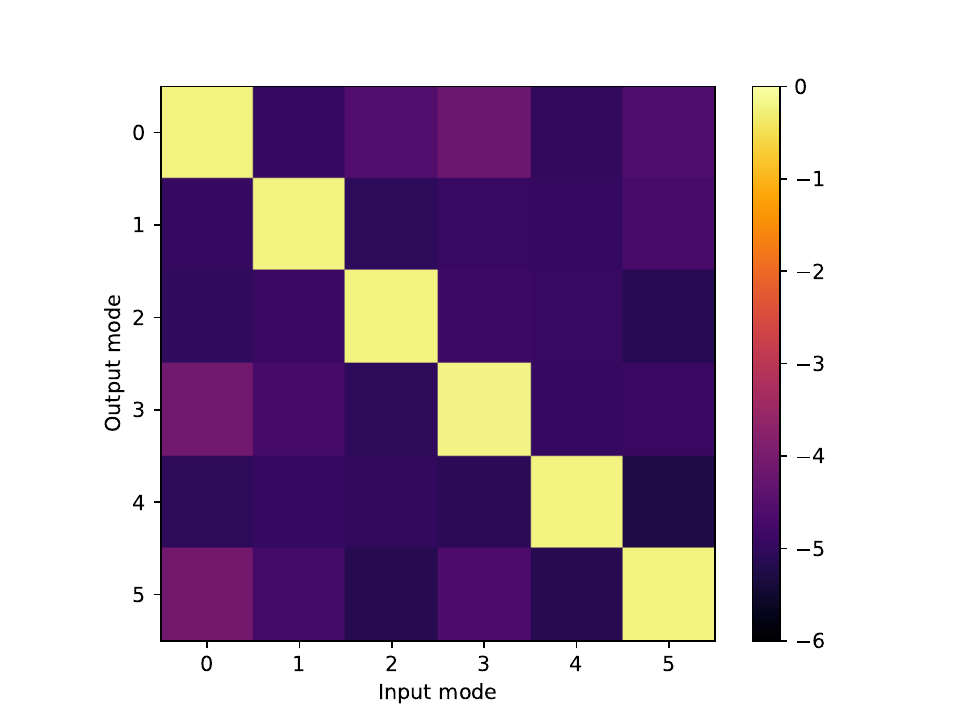}
\centering
\caption{The MPLC SPADE coronagraph sorting the first 6 modes into a hexagonal arangement. Each spot is shaped into a Gaussian spot to further reduce cross-talk caused by diffraction between the modes.}
\label{fig:coronagraph_crosstalk}
\end{figure}

Figure~\ref{fig:coronagraph} shows the simulated performance of an example MPLC SPADE coronagraph. The current design achieves stellar suppression approaching $-50$~dB while maintaining efficient transmission of off-axis light into the desired output modes \ref{fig:coronagraph_crosstalk}. Continued optimization of the phase profiles and the selected modal basis is expected to further improve the achievable contrast. However, the current performance is already beyond the typical cross-talk between modes in photonic lanterns\cite{velazquez2018scaling,becerra2025mode,dana20253d} and what has been achieved with SLM-based MPLCs\cite{deshler2025experimental}.

Unlike conventional coronagraphs, which are generally designed to suppress starlight in the image plane, the SPADE approach directly targets the measurement that maximizes Fisher information for the estimation problem of interest. As a result, the MPLC framework naturally connects coronagraph design with the broader principles of quantum-optimal imaging discussed in this work.

\section{Conclusion and future work}

Spatial mode demultiplexing has recently emerged as a powerful framework for achieving quantum-optimal measurements in astronomy. The principal challenge has shifted from identifying the optimal modal basis to realizing the required optical transformation with sufficient fidelity, bandwidth and scalability. We argue that MPLCs fabricated using patterned liquid-crystal geometric phase technology provide an attractive solution to this challenge.

The three examples presented here demonstrate the breadth of possible applications. Efficient mode matching enables nearly ideal coupling into standard single-mode fibers for segmented telescope pupils, optimized beam combiners achieve the low cross-talk required for future space interferometers, and SPADE-based coronagraphs offer a new route toward quantum-optimal high-contrast imaging. All three instruments are realized using the same underlying MPLC architecture.

Future work will focus on the fabrication and experimental validation of these devices, extension to larger modal bases and the development of complete quantum-optimal instrument concepts for ELT/PCS, GMT/GMagAO-X, HWO and LIFE. We believe that MPLCs represent a promising enabling technology for the next generation of astronomical instruments operating at the fundamental limits imposed by quantum mechanics.

\acknowledgments
S.Y. Haffert, R. Landman, and Y. Xin are supported by NWO award 184.036.004 and VI.Vidi.233.144.

\bibliography{report} 
\bibliographystyle{spiebib} 

\end{document}